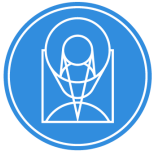

Instrument Science Report WFC3 2025-07

# How Single-Star Guiding Affects HST's Pointing Stability

Jay Anderson & Sylvia Baggett
October 17, 2025

**ABSTRACT**

HST is designed to use two guide stars (GSs) in the fine-guidance sensors (FGSs) to maintain its pointing and tracking during exposures. The primary GS holds the boresight fixed and the secondary GS keeps the orientation fixed. However, HST is also able to track using only a single GS by fixing the boresite on one star and maintaining the orientation using the available gyro(s). We evaluate the pointing quality achieved in this latter case, when one GS and one gyro (RGM, a.k.a., reduced gyro mode) are used. We find that in 1GS-RGM, there is indeed more drift during the course of an orbit than when two guide stars are used, but the drift is much smaller than was seen in previous times of sub-optimal gyro performance. We quantify the 1GS-RGM drift seen in recent archival GO data and in images from a test calibration program and find that (a) for exposures < 500 sec, the PSF quality in 1GS is indistinguishable from that of 2GS and (b) over the course of full orbits (~2500 sec), the drift in four of five cases was ≲ 0.2 pixel and ~0.4 pix in the other case. Such a drift is marginally detectable in observations, but it should have a marginal impact on most science programs, since the variation in the PSF caused by drift is smaller than the PSF variations with focus and with location on the detector. For observers who wish to correct drift effects, we show that the use of a perturbed PSF during post-acquisition data analysis removes essentially all astrometric residuals, even for drift levels up to ~0.5 pix, as well as most of the photometric residuals.

# 1. Introduction

The failure of HST's fourth gyro has left it with only two working gyros, and telescope operations entered reduced gyro mode (RGM) in July 2024. The HST Reduced Gyro Mode Primer (2024) describes the process that led to the adoption of RGM and the limitations that RGM imposes on telescope efficiency and orientation flexibility. As noted in that Primer, HST's pointing stability under RGM is unchanged with respect to its performance under three gyro mode.

The observatory does have some prior experience with RGM limitations. HST operated under RGM from 2005 to 2009, before the gyros were replaced in SM4. Also, from 2021-2024, HST operated under a hybrid three-gyro/RGM mode (see RGM Primer). There are two main consequences of RGM: (1) reduced flexibility in scheduling and (2) somewhat decreased observing efficiency. RGM does not directly change the stability of HST's pointing since HST can use the fine-guidance sensors (FGSs) for high-precision guiding during exposures. However, since RGM can limit the orientations that are possible for a given pointing, there are times when only one guide-star is available for tracking. In addition, even when two guide stars are available, it can be advantageous to use a single GS e.g. to reduce the risk of a losing lock on the stars, which becomes more common as the FGSs age. For these reasons, we assess the pointing quality under 1GS/RGM to determine how it compares to that under 2GS.

The main concern with 1GS is that the telescope uses the single star to hold the boresight fixed in the V2-V3 axis (the telescope focal plane) while control over the V1 axis (i.e., the telescope's orientation) is maintained by the gyro(s). A single gyro is unable to maintain the orientation as well as a second guide star can, thus resulting in some orientation drift. Since the single guide star is held fixed in one of the FGS pickles, the orientation drift results in pointing drift at the location of the detector. **Figure 1** illustrates this with a schematic of the HST focal plane, which has WFC3 at its center. With the guidestar fixed in the V2-V3 (focal) plane, if the orientation changes by the amount indicated by the angle shown by the red arrow, the drift at the detector will be that indicated by the green arrow. The lever arm for WFC3/UVIS is the same for regardless of the FGS containing the GS since WFC3 is at the center of HST's FOV (the lever arm for ACS does depend on which FGS is used for primary-star guiding).

Based on preliminary analysis indicating negligible apparent drifts in 1GS short archival exposures, since Fall 2024 all SNAP program exposures have been acquired under 1GS control. SNAPs are single-orbit visits with typically short exposures. The change improved the SNAP success rate and also provided additional data to quantify any drifts.

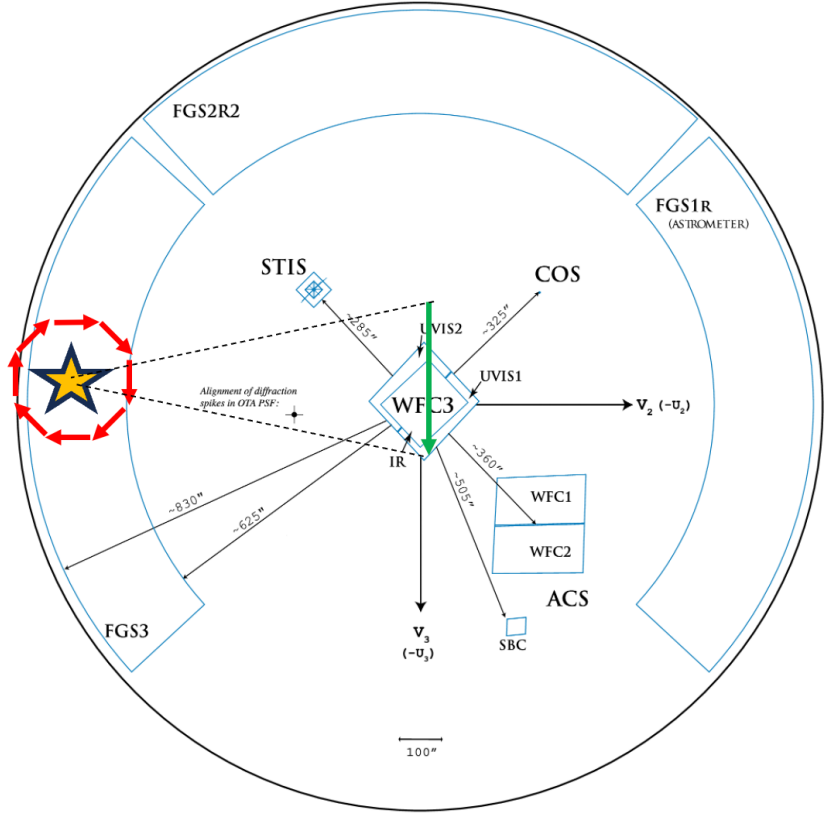

**Figure 1:** Schematic of the HST FOV with a guide star in one FGS. Under 1GS guiding, the telescope rotation about the star (red arrows) is controlled by the gyros, and there can be some roll-angle drift (60″/hour). When the guidestar is far from the science instrument, the drift could cause a boresight shift (green arrow) which would translate into apparent smearing of the target in the image.

The purpose of this document is to evaluate whether 1GS guiding can be effective for most HST programs. We begin by examining the pointing quality in archival data acquired under 1- and 2-GS control. With more than 8 months of SNAP exposures taken under 1GS RGM, we explore its impact on these moderately short exposures in short orbits. Furthermore, we supplemented the SNAP data with any archival GO data that were taken under 1GS/RGM. We analyze all the SNAP+GO data available in **Section 2**, comparing the 1GS/RGM PSF quality to the nominal PSF quality achieved over the lifetime of HST.

While the study of archival data was informative, most exposures were short (<500s). To evaluate the impact of drift on longer exposures, a CAL program (CAL-17893; PI – Anderson) was developed to expand the parameter space. This program obtained images during 5 single-orbit visits to assess the pointing drifts across a full orbit under 1GS/RGM; the results are presented in **Section 3**. In **Section 4**, we investigate how the slight drifting affects the point-spread functions (PSFs), and in turn how such elongated PSFs may impact science. In addition, we show that should drift effects be present, they can be successfully removed in post-acquisition analysis.

## 2. Historical pointing stability

The goal of this document is to determine how pointing/guiding in 1GS/RGM compares to that in 2GS/RGM, as well as how it compares to HST's historic pointing quality.

The information about which and how many guide stars were used in an exposure is available in the _spt engineering files. The _spt header for each exposure contains two relevant keywords: DGESTAR and SGESTAR, which record the names for the dominant and secondary guidestar, respectively. When there is no secondary guide star, the SGESTAR value is set to "none". When DGESTAR and SGESTAR are *both* set to "none", the exposure was a parallel observation. This does not mean that no guide star was used, but simply that the parallel exposure does not have direct access to the guiding information (the information is stored in the prime observations).

Table 1 reports the ten filters with the most WFC3/UVIS exposures taken between May 2021 and November 2024. Also noted is whether a good "library" PSF model exists for that filter (such a model is needed for assessing the PSF quality of an exposure) and how many exposures were taken with that filter under 2GS and 1GS. Finally, the last column in Table 1 indicates whether the filter in question tends to have 5 or more high S/N stars (S/N > 50) in typically observed fields, which is necessary for evaluating the PSF quality. Observations taken with the UV or narrow band filters tend to not have many bright stars unless the target is a globular cluster or a similarly dense field. For this reason, only redder wide-band filters are likely to have suitable exposures for evaluating PSF quality and the image statistics confirm this. WFC3/UVIS images are used, not WFC3/IR, since the WFC3/UVIS pixels are three times smaller (40 mas versus 130 mas) and thus allow a better assessment of any drift.

About half of the usable exposures in Table 1 were taken with F814W, so we focus on this filter since it is more straightforward comparing PSF quality when using the same nominal PSF as a reference.

Table 1: The number of exposures taken in 2-GS and 1-GS mode for each of the top ten most popular WFC3/UVIS filters between May 2021 and November 2024. The filters with available PSF models as well as exposures containing a suitable number of stars in a typical image are highlighted in green.

| FILTER | PSF? | 2-GS | 1-GS | S/N>50 Stars? |
|---|---|---|---|---|
| **F814W** | **Y** | **4830** | **349** | **Y** |
| F275W | Y | 3569 | 131 | N |
| **F606W** | **Y** | **3108** | **175** | **Y** |
| F336W | Y | 2389 | 62 | N |
| F350LP | N | 2361 | 38 | N |
| F656N | N | 2171 | 17 | N |
| **F555W** | **Y** | **1717** | **74** | **Y** |
| F475W | N | 979 | 104 | Y |
| F225W | Y | 974 | 34 | N |
| F438W | Y | 702 | 20 | N |

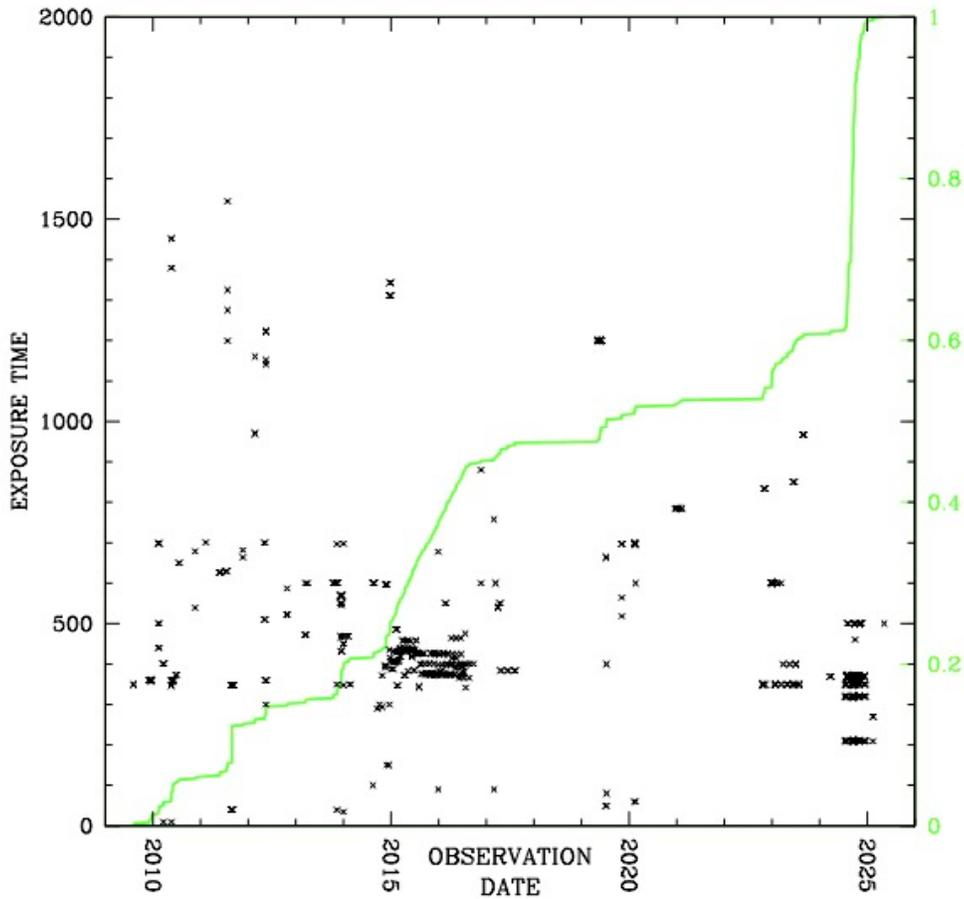

**Figure 2: All exposures taken with one guide star through F814W over the lifetime of WFC3/UVIS. The cumulative distribution is in green.**

We identified all F814W exposures taken over the lifetime of WFC3 and characterized them as 1GS or 2GS and 3GM or RGM. **Figure 2** plots the exposure time of each 1GS exposure as a function of observation date since 2009; the green line shows the cumulative distribution of 1GS exposures over time. During most years of WFC3 operation, it was, as expected, unusual for images to be acquired with just 1GS. Most of the excess of points between 2015 and 2016 were from two related programs (GO-13677 and GO-14327) that prioritized scheduling over 2GS operations. The jump in the number of short (500s) exposures after 2024 is due to the recent requirement of 1-star guiding for SNAP programs.

In order to assess the PSF quality in each image, we ran the `hst1pass` routine (Anderson 2022) on each of the ~5000 1GS and 2GS exposures taken with F814W. We used the "library" nominal PSF file, STDPSF_WFC3UV_F814W.fits[1], rather than the focus-diverse version. Although it is well known that the shape and sharpness of HST's PSF can change due to breathing-related focus changes, the trailing we are looking to evaluate could be mistaken for breathing-related defocus, thus we used the nominal optimal-focus PSF. Breathing will indeed introduce variations in the PSF fit, but these variations should simply introduce some spread about the average, and this spread should be the same for 1GS and 2GS images.

---

[1] https://www.stsci.edu/hst/instrumentation/wfc3/data-analysis/psf

The `hst1pass` routine produces two main PSF-fit-related diagnostics for each star that it fits. The most commonly used diagnostic is `qfit`, which reports for each measured star the absolute total residuals of the PSF fit to the star's inner 5×5 pixels, divided by the flux of the star. Defined this way, `qfit` has a value between 0 and 1. Stars that are well fit by the model PSF tend to have `qfit` values less than 0.05. The `qfit` metric is more useful than, say, chi-squared, when the error in the fit comes from the PSF rather than the source, since the `qfit` metric better tracks PSF errors independent of the noise in the star (provided that the star has S/N ≥ 50).

Another metric produced by `hst1pass` is the excess/deficit flux in the star's central pixel relative to the PSF-model prediction. This is reported as column "C" in the output file (note the capital case; lowercase "c" refers to chi-squared). Like q, C is also normalized to the flux of the star: C = $(P_{IJ} - s_* - z_* \psi_{IJ})/z_*$, where the star's central pixel is [I,J], and $P_{IJ}$ is the flux in that pixel, $s_*$ is the sky value, $z_*$ is the solved-for flux of the star, and $\psi_{IJ}$ is the fraction of the star's flux that should have landed in the star's central pixel according to the PSF model. When the PSF is a good fit to the star, there is no systematic residual at its center, and C is distributed about zero in accordance with the Poisson noise of the star. However, when the star image is broader than the model PSF, there is a deficit of flux in the central pixel, and C for that star image is negative. If the detection is a cosmic ray, which are usually sharp relative to the PSF, then C is positive. We ran the `hst1pass` routine on all 5000 images taken with F814W over the 15-year time period and selected those images that had at least ten stars with good S/N[2].

Before we can evaluate the impact that 1-star guiding has on PSF quality, we must first get a sense of what the "typical" PSF quality is. Since small guiding issues have a larger effect on the central pixel intensity of the PSF more than on the PSF fit over the inner 5x5 pixels, we focus here on the C parameter reported by `hst1pass`.

**Figure 3** presents various metrics for two-star guiding on the left and for one-star guiding on the right. The top row shows the RMS of the guide-star pointing as a function of exposure time from the engineering data in the _spt files. The typical 2GS RMS jitter is about 5 mas (0.125 pixel), but there is a sharp peak in jitter up to about 15 mas (0.375 pixel) for observations taken between 2018.3 and 2018.8 (colored in blue). Those jitter excursions were related to a failing gyro which was taken out of service in April 2018. In the upper right panel, we see that the 1GS observations appear to have nominal guiding quality for essentially all the observations. No 1GS observations were taken during 2018 when the gyro was failing.

The bottom plots show the central PSF residual metric C averaged over all the good S/N stars in each exposure. Each point corresponds to the average central PSF residual for one exposure, and only exposures with at least 10 good S/N stars are shown. The exposures taken while one of the gyros was failing (the blue points) clearly exhibit larger central-pixel deficits related to broadening of the PSF, with the central pixel low by about 1.5%. For context, a star that is centered on a pixel typically receives about 16% of its F814W light in its central pixel thus the guiding-impacted stars received about 14.5% due to the additional PSF broadening. (We will quantify the expected broadening as a function of jitter in **Section 4**.) The recent exposures taken in RGM (green points), both 1GS and 2GS, exhibit much smaller PSF deficits at their centers.

---

[2] Stars must be unsaturated and have an instrumental magnitude brighter than –9, i.e. at least 4000 counts total within a 10-pixel radius.

We note that the 1GS data taken under three-gyro mode (the black points in the plots on the right) are qualitatively different from the 1GS data taken during RGM (the green points). With the latter, we would expect the pointing to drift along the green path in **Figure 1** as the telescope V1 axis rotates about the guide star due to the slow gyro drift, even as the pointing control loop keeps the guide star nailed at its location in the FGS. The black points show that even with one guide star, the three gyros were generally able to maintain good orientation control. The few large-residual 1GS values were mainly from observations between 2017.0 and 2019.5 when the gyros were becoming unreliable.

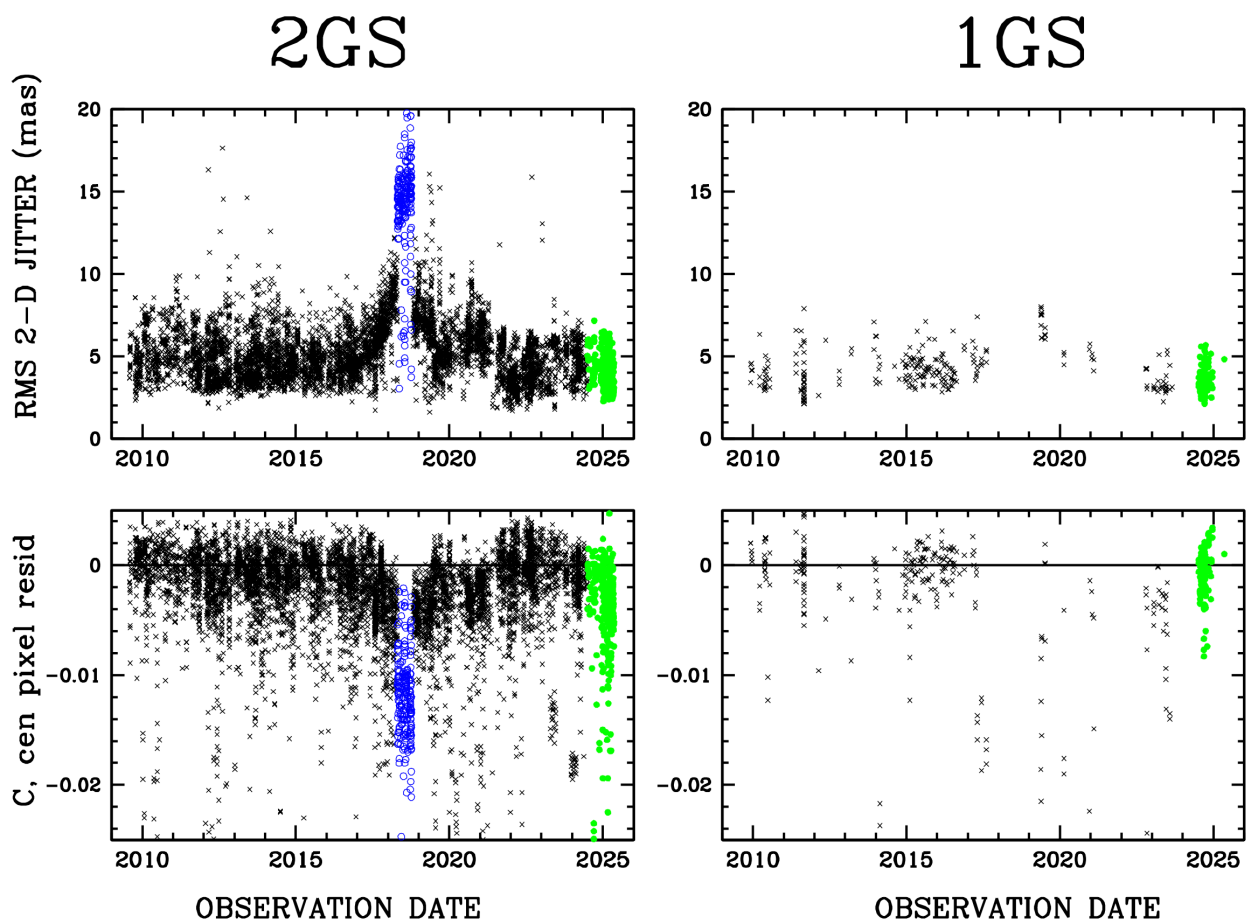

Figure 3: Distribution of jitter (top) and central PSF excess (bottom) for F814W observations taken with two guide-stars (left) and one guide-star (right). Each point corresponds to one exposure, and each exposure had at least 10 well-measured stars (to ensure robust PSF analysis). The top row shows the distribution of jitter (from engineering data) as a function of observation date. The blue points correspond to observations between 2018.3 and 2018.8, when one of the gyros was becoming unreliable and increasing the jitter. The exposures in green were taken after 2024.5, when RGM was activated. The bottom row of plots shows the `hst1pass` quantity C, the central-pixel residual relative to the PSF, as a function of observation date.

The PSF quality as a function of exposure time for 2GS observations and for 1GS observations under three-gyro control showed no trend. As described earlier, the 1GS/RGM exposures were almost all associated with SNAP programs, so they didn't probe exposures longer than about 500s. The lack of 1GS+RGM observations with exposure times longer than 500s makes it difficult to formulate conclusions concerning pointing quality in longer 1GS/RGM exposures (e.g. full-orbit 2000-s exposures). As a result, a calibration proposal (CAL-17893, PI-Anderson) was developed to acquire data with which to address this and is presented in the next section.

## 3. CAL-17893

Program CAL-17893 was developed to acquire observations to allow an evaluation of how much boresight drift WFC3 experiences over the course of full orbits under 1GS/RGM . The program consisted of five single-orbit visits where each visit contained one short 35-s exposure and five 350s exposures. No dithering was used, ensuring that any difference in telescope pointing could be attributed to guiding issues. The goal was for the five different visits to provide five separate realizations of the drift over the course of a full orbit.

CAL-17893 targeted IC 1848, a moderate density young open cluster that had observability in the February 2025 timeframe, observing a region just south of the cluster center to avoid two bright blue stars. Since the filter choice has no impact on measuring pointing shifts, we chose the F410M filter which is used for phase-retrieval measurements (Dressel & Rivera 2024), thereby allowing the images to serve two calibration purposes. There were about 60 good S/N>50 stars in each of the 350s WFC3/UVIS exposures.

The first visit was taken on 11 Feb 2025. All 6 exposures were acquired, but they were beset by guiding issues: there were streaks in some images and no point-like stars. The second visit was taken the next day on 12 Feb 2025 and suffered similar streaking. Visits 3, 4 and 5 executed without issue on 14, 16, and 17 Feb 2025. Two orbits were granted to repeat visits 1 and 2; these (visits 51,52) were successfully executed on 30 Mar 2025. Table 2 lists all the successful exposures.

We ran `hst1pass` on the 30 good exposures using a nominal F410M PSF model, extracting a list of good S/N stars, and determining `q` and `C` metrics for each star in each exposure. We collated the good S/N stars in each exposure for each visit with the stars in the first long exposure for that visit and determined the boresight offset. That offset is reported as $\Delta x_1$ and $\Delta y_1$ in the table, in units of UVIS pixel size. By construction, the offset for the first long exposure is zero. By combining the five long exposures in various ways, we can simulate exposures of different quantized lengths and assess the pointing drift during an orbit. For example, combining the five long exposures together simulates the drift experienced by a full-orbit exposure of about 40 minutes. In this case, the "average" pointing is not the pointing of the first exposure, but rather the average offset of the five long exposures. The offset of each exposure relative to this average pointing is reported as $\Delta x_B$ and $\Delta y_B$ in the table. The RMS drift during the exposure is the RMS relative to this average.

**Table 2:** Successful observations for CAL-17893 with measured drift offsets relative to the first long exposure ($\Delta x_1$ and $\Delta y_1$) and relative to the average pointing of the five long exposures (($\Delta x_B$ and $\Delta y_B$).

| IMAGE NAME | DATE | TIME | $\Delta T$ (min) | PA_V3 (°) | FILTER | EXPT (s) | $\Delta x_1$ (pix) | $\Delta y_1$ (pix) | $\Delta x_B$ (pix) | $\Delta y_B$ (pix) |
|---|---|---|---|---|---|---|---|---|---|---|
| **ifjl03srq** | 2/14 | 00:59 | −2 | 250.85 | F410M | 35  | +0.10 | +0.09 |       |       |
| **ifjl03ssq** | 2/14 | 01:01 | 0  | 250.85 | F410M | 350 | —     | —     | +0.22 | +0.21 |
| **ifjl03suq** | 2/14 | 01:09 | 8  | 250.85 | F410M | 350 | −0.13 | −0.12 | +0.09 | +0.09 |
| **ifjl03swq** | 2/14 | 01:17 | 8  | 250.85 | F410M | 350 | −0.27 | −0.22 | −0.05 | −0.01 |
| **ifjl03syq** | 2/14 | 01:25 | 8  | 250.85 | F410M | 350 | −0.34 | −0.31 | −0.12 | −0.10 |
| **ifjl03t1q** | 2/14 | 01:33 | 8  | 250.85 | F410M | 350 | −0.35 | −0.42 | −0.13 | −0.21 |
| **ifjl04csq** | 2/16 | 01:31 | −3 | 249.49 | F410M | 35  | −0.05 | −0.03 |       |       |
| **ifjl04ctq** | 2/16 | 01:34 | 0  | 249.49 | F410M | 350 | —     | —     | +0.07 | +0.01 |
| **ifjl04cvq** | 2/16 | 01:42 | 8  | 249.49 | F410M | 350 | +0.03 | +0.05 | +0.10 | +0.06 |
| **ifjl04cxq** | 2/16 | 01:50 | 8  | 249.49 | F410M | 350 | −0.03 | +0.01 | +0.04 | +0.02 |
| **ifjl04czq** | 2/16 | 01:58 | 8  | 249.49 | F410M | 350 | −0.12 | −0.03 | −0.05 | −0.02 |
| **ifjl04d2q** | 2/16 | 02:06 | 8  | 249.49 | F410M | 350 | −0.21 | −0.07 | −0.14 | −0.06 |
| **ifjl05duq** | 2/17 | 21:31 | −3 | 248.26 | F410M | 35  | −0.06 | −0.02 |       |       |
| **ifjl05dvq** | 2/17 | 21:34 | 0  | 248.26 | F410M | 350 | —     | —     | −0.00 | −0.05 |
| **ifjl05dxq** | 2/17 | 21:42 | 8  | 248.26 | F410M | 350 | +0.08 | +0.09 | +0.08 | +0.04 |
| **ifjl05dzq** | 2/17 | 21:50 | 8  | 248.26 | F410M | 350 | +0.07 | +0.10 | +0.07 | +0.05 |
| **ifjl05e1q** | 2/17 | 21:58 | 8  | 248.26 | F410M | 350 | −0.01 | +0.04 | −0.01 | −0.01 |
| **ifjl05e4q** | 2/17 | 22:06 | 8  | 248.26 | F410M | 350 | −0.12 | +0.02 | −0.12 | −0.03 |
| **ifjl51ggq** | 3/30 | 04:32 | −3 | 219.29 | F410M | 35  | +0.09 | -0.00 |       |       |
| **ifjl51ghq** | 3/30 | 04:35 | 0  | 219.29 | F410M | 350 | —     | —     | +0.18 | −0.01 |
| **ifjl51gjq** | 3/30 | 04:43 | 8  | 219.29 | F410M | 350 | −0.14 | +0.01 | +0.04 | +0.00 |
| **ifjl51glq** | 3/30 | 04:51 | 8  | 219.29 | F410M | 350 | −0.24 | +0.01 | −0.06 | +0.00 |
| **ifjl51gnq** | 3/30 | 04:59 | 8  | 219.29 | F410M | 350 | −0.27 | +0.01 | −0.09 | +0.00 |
| **ifjl51gqq** | 3/30 | 05:07 | 8  | 219.29 | F410M | 350 | −0.26 | −0.00 | −0.08 | −0.01 |
| **ifjl52hiq** | 3/30 | 06:07 | −2 | 219.23 | F410M | 35  | +0.05 | −0.01 |       |       |
| **ifjl52hjq** | 3/30 | 06:09 | 8  | 219.23 | F410M | 350 | —     | —     | +0.09 | −0.01 |
| **ifjl52hlq** | 3/30 | 06:17 | 8  | 219.23 | F410M | 350 | −0.08 | +0.01 | +0.01 | +0.00 |
| **ifjl52hnq** | 3/30 | 06:25 | 8  | 219.23 | F410M | 350 | −0.15 | +0.02 | −0.06 | +0.01 |
| **ifjl52hpq** | 3/30 | 06:33 | 8  | 219.23 | F410M | 350 | −0.14 | +0.01 | −0.05 | +0.00 |
| **ifjl52hsq** | 3/30 | 06:42 | 9  | 219.23 | F410M | 350 | −0.09 | −0.01 |  0.00 | −0.01 |

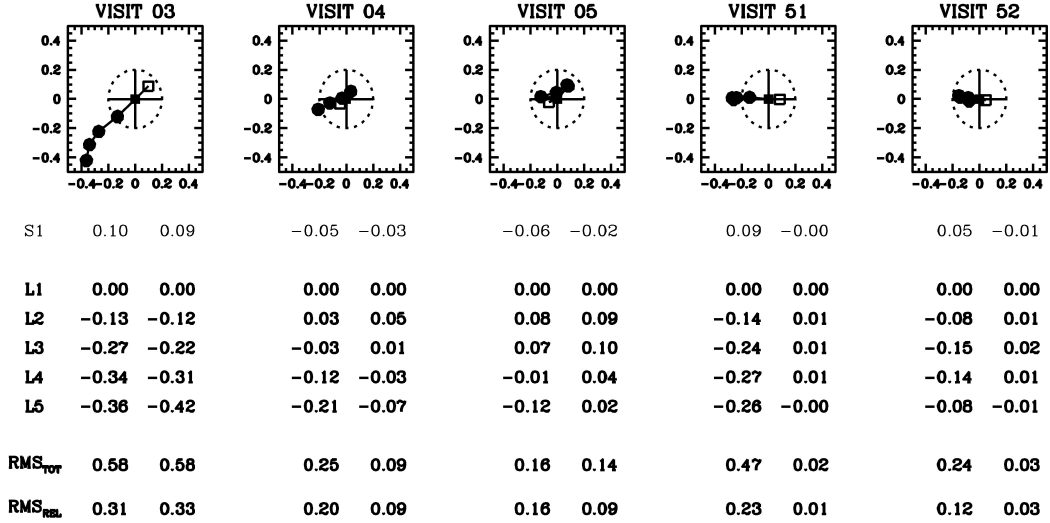

**Figure 4:** Drift results for exposures from CAL-17893. The plots show the offset in WFC3/UVIS pixels for the exposures in each visit relative to the pointing of the first long exposure. The short exposure at the beginning of the orbit is shown as an open symbol. The table below the figure provides the offsets for the exposures (S1 through L5) along with the RMS relative to the first exposure and the RMS relative to the average.

The data in **Table 2** are shown graphically in **Figure 4**. There is quite a bit of drift in visit 3, but the other four visits show marginal drifts, which is reflected in the RMS values reported in the table at the bottom of the figure as well. It is worth noting that the PSF quality in the images of the visit 3 is similar to that of the others visits; apart from the drift between exposures, the observations are otherwise nominal. $RMS_{TOT}$ provides the total RMS drift across the orbit relative to the initial pointing, while $RMS_{REL}$ shows the RMS relative to the average pointing. This second value is more relevant when we consider any blurring in the PSF for a long exposure. Four of the five visits show total relative drifts of 0.22, 0.18, 0.23, and 0.12 pixel; the worst visit shows a total relative drift of 0.45 pixel (0.31 + 0.33 in quadrature).

**Figure 5** shows the location of the guide stars for the five visits relative to the detector. The WFC3/UVIS FOV is outlined in blue while the other slightly larger square FOV is ACS/WFC. In the first three successful visits (3, 4, and 5), the roll drift would be at about 30° to the left below the WFC3 $\hat{x}$ axis, which is what we see, and for the last two visits (51 and 52) roll drift would be mostly along the WFC3 $\hat{x}$ axis.

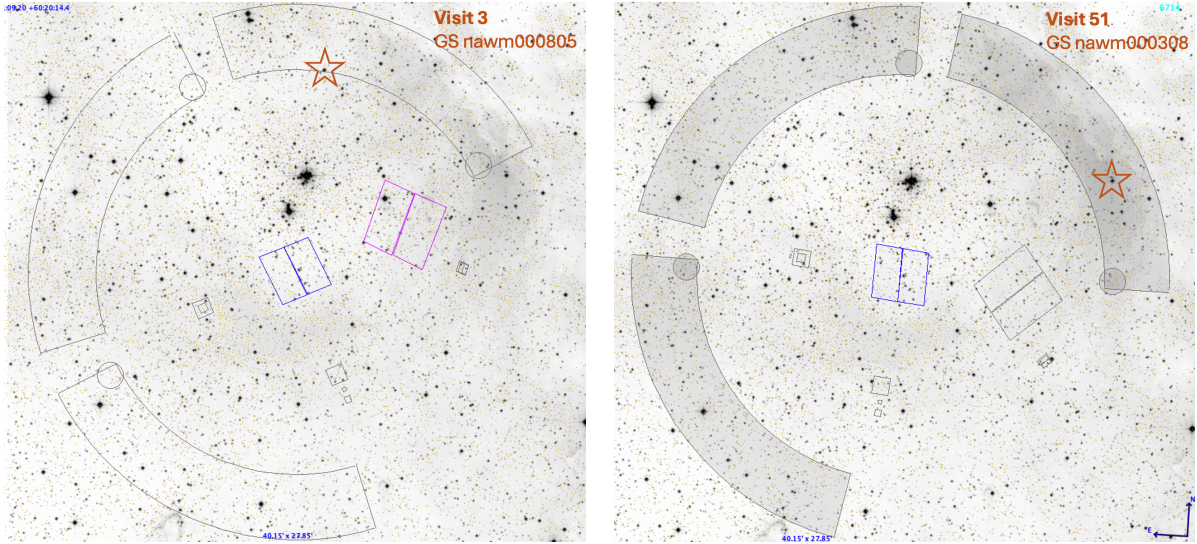

**Figure 5:** Schematic showing the location of the guide stars (orange) used for visits 3, 4, and 5 (left) and visits 51 and 52 (right) relative to the UVIS detector (blue).

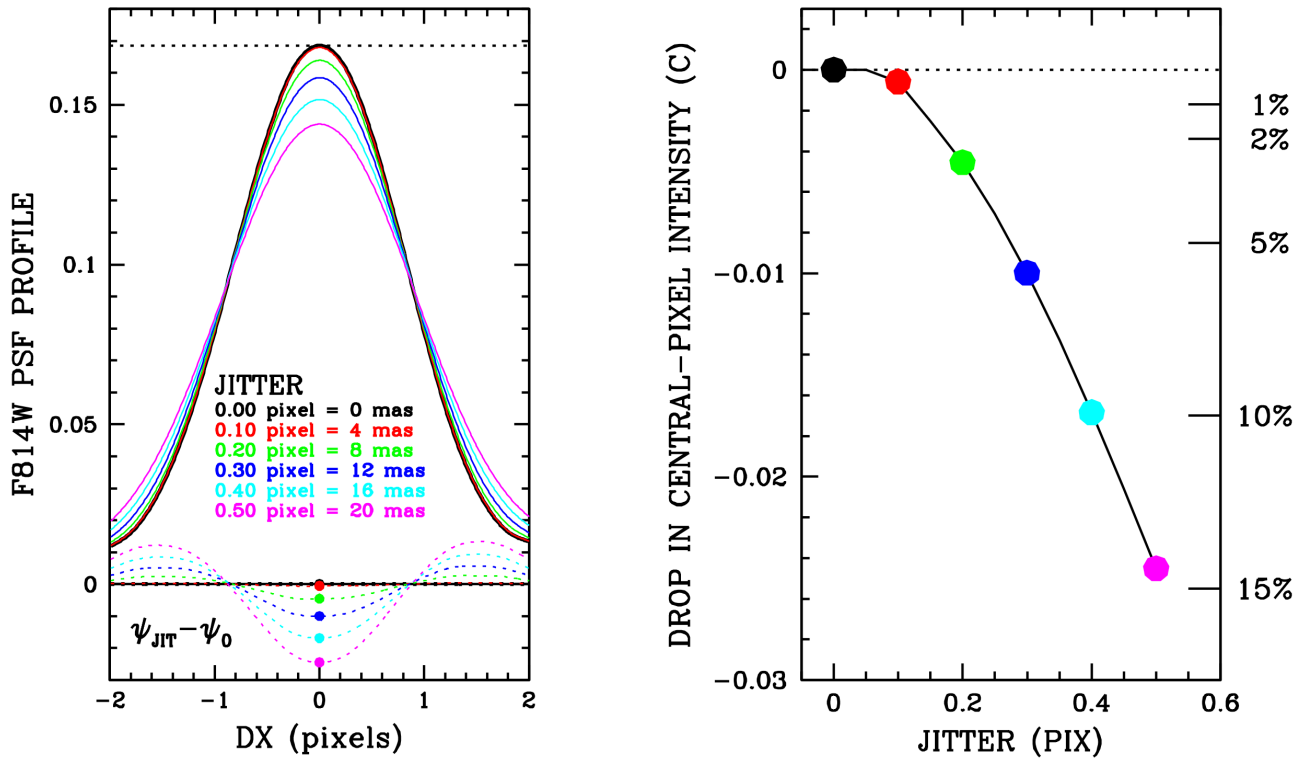

**Figure 6:** (Left) Profile of the effective PSF for WFC3/UVIS's F814W filter. Black curve corresponds to nominal jitter, and the other five curves show the PSF after the labeled amounts of 1-D jitter in $\hat{x}$ is convolved with the PSF. (Right) The drop in central pixel flux as a function of jitter.

# 4. Impact of Guiding Drifts on Science

In **Section 2**, we studied the historical guiding quality in terms of how guiding affects the fraction of light that lands in a star's central pixel. **Figure 3** shows that guiding for HST tends to be extremely stable: the fraction of light that lands in the central pixel of a star generally varies by less than 0.3% in an absolute sense (which is 2% in a relative sense, namely 0.003 over 0.15, the baseline fraction received by the central pixel). Only during periods of significant gyro instability did this approach 1.2% absolute (8% relative) and, even then, most HST science was achievable without any special analysis tools.

The values in **Figure 3** provide a sense of how stable the HST PSF is, but they do not provide a direct measure of how the variations in the PSF may impact science. To address this, we performed simulations using actual HST PSFs. We started with the "library" effective PSF for F814W at the center of WFC3/UVIS's bottom detector (UVIS2). The heavy black curve in **Figure 6** shows a horizontal slice through the center of this PSF: a star centered on a pixel will have 16.8% of its light in its centermost pixel. We convolved this library PSF with a range of 1-D jitter along $\hat{x}$, at levels from 0.1 pixel to 0.5 pixel RMS. The horizontal slices through the centers of the resulting PSFs are shown in the left panel with different colors. The black curve corresponds to nominal jitter, and the other curves show progressively wider profiles and progressively less flux in the central pixel. The dotted lines at the bottom show the difference between the nominal-jitter PSF and the added-jitter PSFs. A jitter of 0.2 pixel decreases the fraction of light in the central pixel from 16.8% to 16.5%, which is an absolute change of 0.3% and a relative change of about 2%.

The 0.2-pixel elongation corresponds to the nominal separation between two stars that can be reliably distinguished as a double star with HST (assuming good S/N). The right panel of **Figure 6** shows the central-pixel deficit as a function of jitter (the "C" parameter from `hst1pass`).

These variations in the central pixel flux C can be placed into context by examining how much this quantity varies compared to other well-known sources of PSF variation such as location on the detector and with focus. The left panel of **Figure 7** shows, as in **Figure 6**, the horizontal profile through the center of F814W PSFs at a variety of one-dimensional RMS drifts.

The middle panel shows the same profile, but for the nominal PSF at a regular array of locations across the detector. The central pixel receives $16.8 \pm 0.08$ percent of a star's light. The RMS variation with location on the detector is about 5% about the average, equivalent to about 0.25 pixel of 1-D RMS jitter relative to nominal (reading off from the right panel of **Figure 6**).

The rightmost panel of **Figure 7** shows how the PSF varies with focus. We plot the same horizontal slice of the F814W PSF for the ten different focus levels studied in Anderson (2018). *Figure 7 in that study shows that 75% of observations have focus levels that lie in the middle four focus zones, so the typical variation is probably half of this. However, observers are not guaranteed images with any particular focus, so the full range of focus values are possible for any observation.*

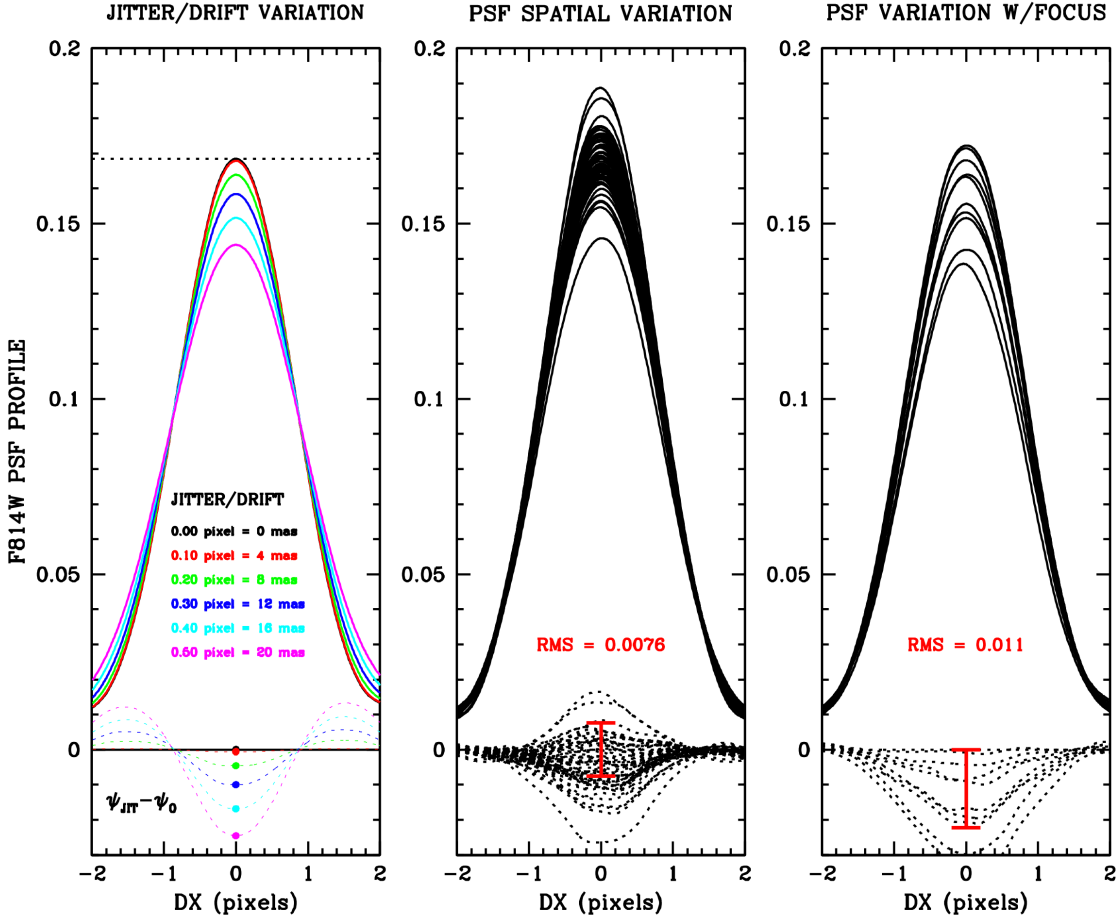

**Figure 7:** (Left, repeated from Figure 6) Profile of the F814W effective PSF with a variety of added 1-D jitter. (Middle) Profile of the F814W effective PSF shown at an array of locations across the detector. (Right) Profile of the F814W effective PSF at 10 different focus levels.

## 5. Analysis with slightly drifted images

While the impact of a slight drift on images may induce variations in the PSF that are similar in amplitude to variations typically seen with position and focus (for which PSF models exist), the drift-impacted variations must be handled separately. In this section, we simulate trailed observations with the F814W PSF and evaluate how the trailing impacts the ability to measure photometry and astrometry of stars.

We generated an array of 100×100 star images spanning the detector, separated by ~40 pixels, and sampling every 0.01 pixel of pixel phase in x and y. These 10,000 stars were inserted with an instrumental magnitude of −12.5 (a total of 100,000 counts) and with the nominal F814W PSF at the center of the bottom chip. No spatial variation is contained in the PSFs. In the first simulated image, we inserted the PSF as-is, with no trailing, and in the subsequent images, we convolved the PSF with Gaussians that had RMSs of 0.1 pixel, 0.2 pixel, 0.3 pixel, 0.4 pixel, and 0.5 pixel, all in $\hat{x}$. We also simulated trailing aligned at 30 degrees relative to $\hat{x}$. **Figure 8** shows a 200×100 pixel region of the initial untrailed artificial image.

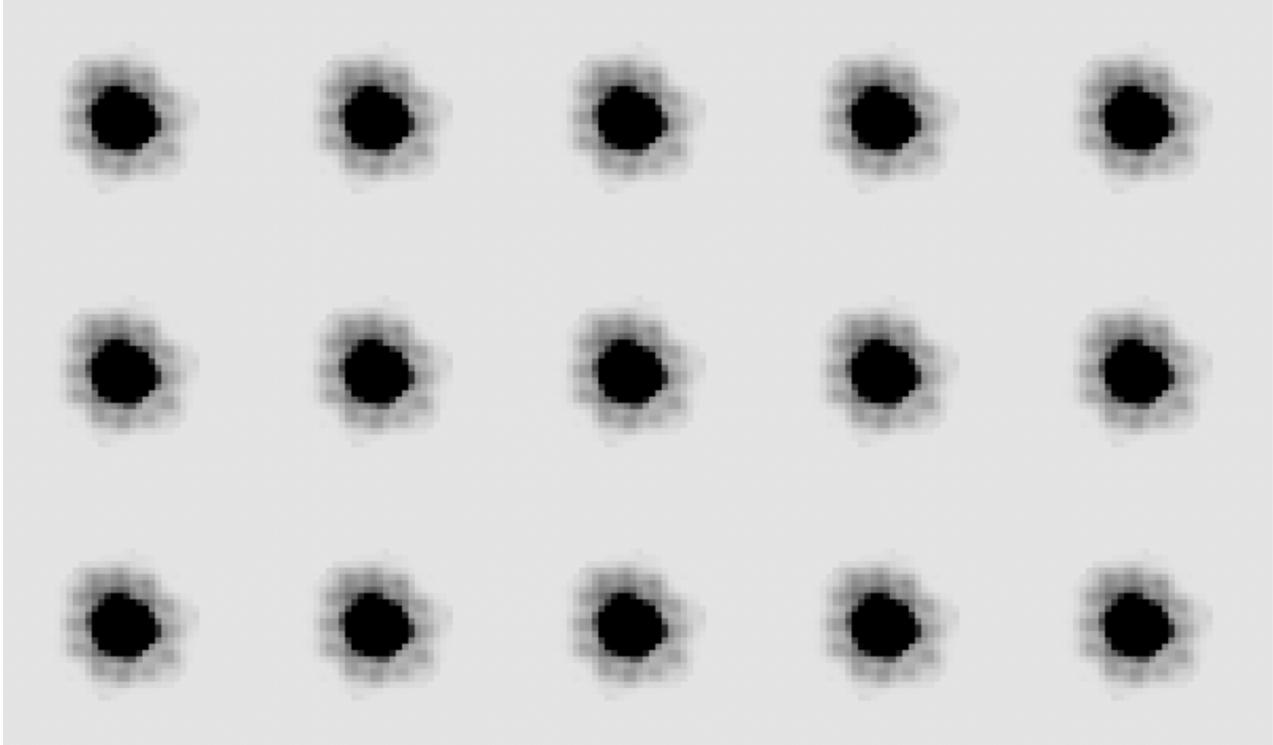

Figure 8: A 200x100-pixel region of the added untrailed stars, which spanned all pixel phases.

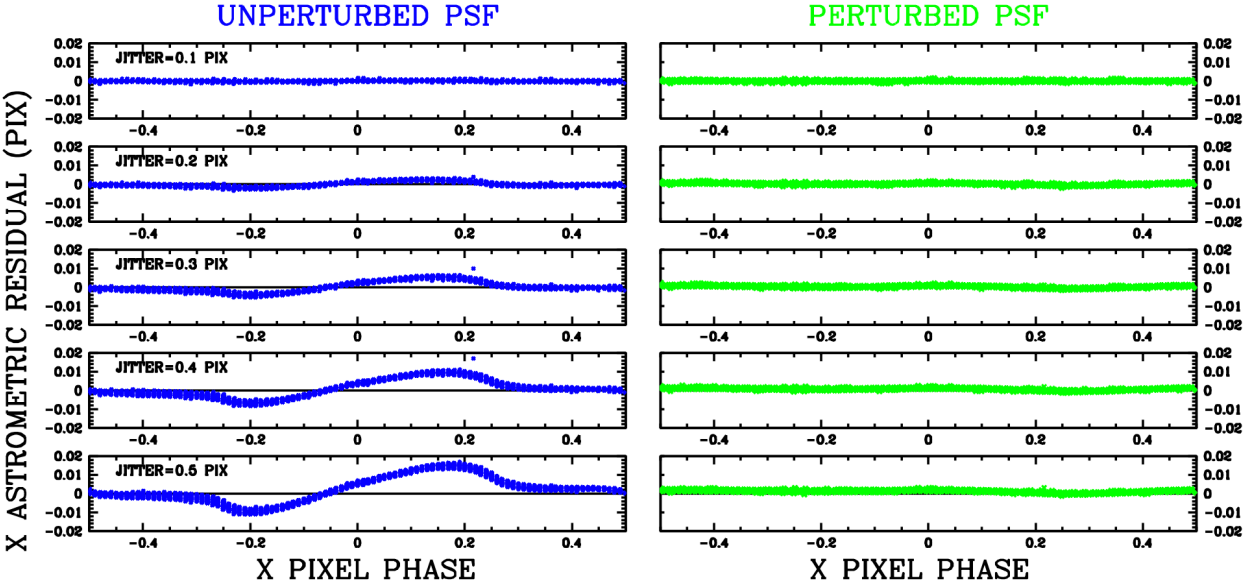

Figure 9: The astrometric residuals as a function of pixel phase for simulated trailed observations, with the effective 1-D jitter along x as labeled within the plot. Residuals measured after analysis with the nominal PSF and a perturbed PSF are shown at left and right, respectively.

We analyzed this set of images in two different ways. For the first approach, we ran `hst1pass` on each image using the spatially constant PSF that was the basis for the un-trailed simulations and produced a list of measured positions and fluxes for each star. We collated the list for each trailed exposure against the list for the untrailed exposure and examined the residuals, finding a familiar issue. The left panels of **Figure 9** show the astrometric residuals in the $\hat{x}$ direction as a function of $\hat{x}$ pixel phase, $x - \text{int}(x+0.5)$. Basically, the residuals reflect what happens when you measure a star with a PSF that is too broad (or narrow). When the star is centered on the pixel, it has equal fluxes in the left and right pixels, so the width of the PSF is irrelevant to the astrometry. Similarly, when a star is centered on the boundary between two pixels, the flux is equal between these two pixels and the width of the PSF is irrelevant. But if a star is somewhere between the center of a pixel and the edge of a pixel, the PSF model is needed to specify how flux transfers from one pixel to the other as the star's position shifts. This non-linear transfer of flux is faster when the intrinsic PSF is sharper, and errors in the model-versus-actual PSF show up as the sine curve above (see the discussion of "pixel-phase errors" in Anderson & King 2000).

The right panels of **Figure 9** show that if we allow `hst1pass` to construct a "perturbation" to the library PSF, then the residuals effectively disappear, even when the added jitter is high (0.5 pixel). The PSF perturbation itself is shown in **Figure 10**. The decrement in the fraction of light in the central pixel is shown for each simulation. Since 16.8% of a star's light lands in its central pixel when it is centered on that pixel, the drop of 0.016 in the 0.4-pixel-drift perturbation means that 10% less light lands in the central pixel. This light is redistributed to the pixels to the right and left.

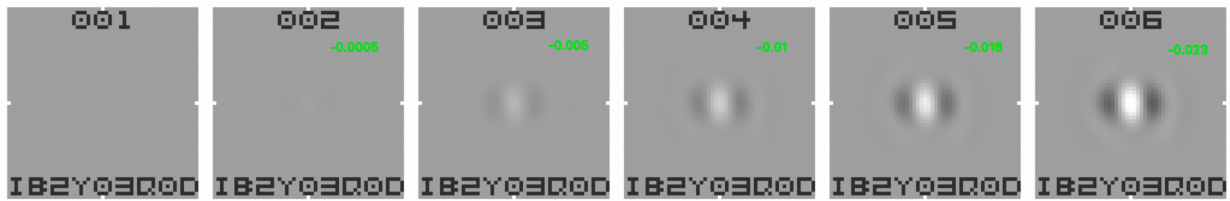

**Figure 10: The perturbation PSF for the nominal image (left) and the images with 0.10, 0.20, 0.30, 0.40, and 0.50 pixels of jitter in the x direction. The decrement in central pixel flux (in terms of fractions of the total PSF) are shown for each in green. Black signifies more flux and white less flux.**

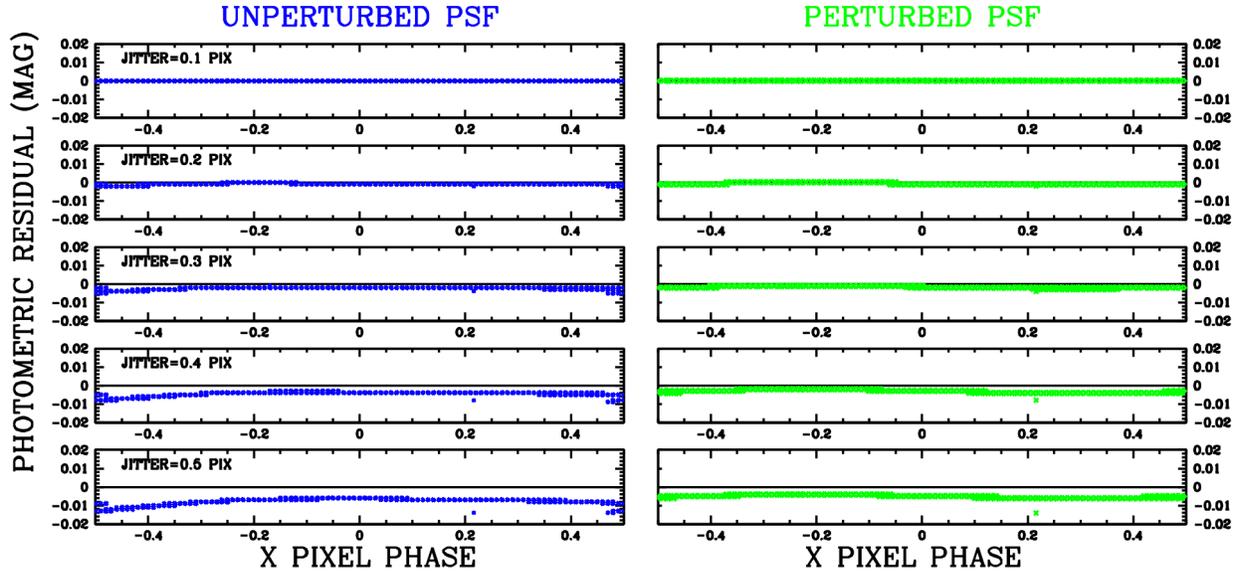

Figure 11: The photometric residuals as a function of pixel phase for simulated trailed observations, with the effective 1-D jitter along x as labeled. Residuals measured after analysis with the nominal PSF and a perturbed PSF are shown at left and right, respectively.

.

In **Figure 11**, we examine the photometric residuals as a function of pixel phase. We find that these residuals show a different behavior from the astrometry residuals. Naturally, any jitter or trailing will have a smaller effect on photometry than astrometry, since astrometry is focused on differences of pixel values within the inner 5×5 pixels of a stars, and photometry is focused only on the sum of the pixel values. Any effect that moves flux from the central pixel to the immediately surrounding pixels will not affect aperture-based photometry. That said, the 0.5-pixel jitter does end up sloshing about 1% of the flux out of the 5×5-pixel measurement aperture relative to the nominal PSF. The `hst1pass` perturbation PSF analysis can reduce the photometry residuals by about half, but the `hst1pass` perturbation is focused on the inner shape rather than the total flux, so the correction is not perfect. *Nevertheless, if one measures relative fluxes with `hst1pass` and then uses a 5- to 10-pixel aperture on the `_drz` images to perform the absolute calibration (as is recommended), then even this small residual will disappear, and the jitter will have negligible impact on the large-aperture flux.*

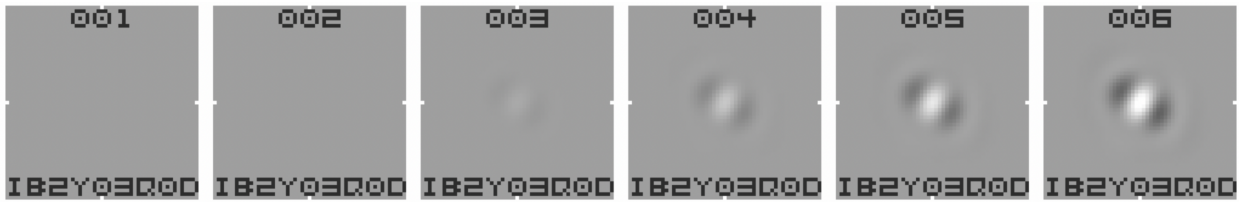

Figure 12: Perturbation PSF for images that were trailed along a direction that is 30° inclined relative to the  degrees.

We performed similar perturbed and unperturbed comparisons for images that were trailed in a direction that was 30° inclined relative to the $\hat{x}$ axis. **Figure 12** shows the perturbation PSF extracted for this simulation. The results were the same as those for the jitter aligned along the $\hat{x}$ axis: the perturbation PSF does an excellent job restoring unbiased astrometry.

## 6. Conclusions

This document presents an assessment of the guiding quality that can be achieved with 1GS under RGM. All available archival exposures taken with a single guide star through F814W were evaluated: the PSFs in these 1GS exposures are indistinguishable in quality to the PSFs in exposures taken in 2GS. Since most of these archival exposures were SNAP exposures, which tend to be short (< 500 s), a five-orbit calibration program (CAL-17893) was executed to provide data to probe the impact of gyro drift during long exposures. Four orbits exhibited drifts of less than 0.25 pixel total from beginning to end (net RMSs of about 0.15 pixel per coordinate) while one orbit exhibited a drift of almost 0.5 pixel total from beginning to end (a net RMS 0.3 pixel per coordinate). The quality of 1Gs imaging will continue to be closely monitored.

The nominal guiding quality is ~7 mas per coordinate axis (0.18 UVIS pixel). Since an additional drift-related RMS would more than double the jitter along one axis, we evaluated the impact of additional PSF broadening on HST's science. We simulated drift-broadening of the PSF by 0.1, 0.2, 0.3, 0.4, and 0.5 pixel in RMS, both along a coordinate axis and at an angle to the coordinate axes. Broadening of 0.2 pixel reduced the fraction of light in the core from about 16.8 to 16.0%, while a broadening of 0.5 pixel reduced it to about 14.5%. *Even with this broadening, however, it was possible to do systematically accurate astrometry and photometry by perturbing the empirical model PSFs to better match the PSFs of stars in each individual image.*

### *Acknowledgements*

We thank WFC3 team reviewers Amanda Pagul and Joel Green for their comments and suggested improvements to this report.